\documentclass[fleqn,usenatbib]{mnras}

\usepackage{newtxtext,newtxmath}
\usepackage[dvipsnames]{xcolor}
\usepackage[T1]{fontenc}
\usepackage{array}
\DeclareRobustCommand{\VAN}[3]{#2}
\let\VANthebibliography\thebibliography
\def\thebibliography{\DeclareRobustCommand{\VAN}[3]{##3}\VANthebibliography}

\usepackage{graphicx}	% Including figure files
\usepackage{amsmath}	% Advanced maths commands
\usepackage{hyperref}
\title[Probing cosmic homogeneity in the Local Universe]{Probing cosmic homogeneity in the Local Universe}

\author[B. L. Dias et al.]{
Bruno L. Dias,$^{1}$\thanks{E-mail: brunoleal@on.br}
Felipe Avila,$^{1}$
Armando Bernui$^{1}$
\\
$^{1}$Observat\'orio Nacional, Rua General Jos\'e Cristino 77, 
	S\~ao Crist\'ov\~ao, 20921-400 Rio de Janeiro, RJ, Brazil
}

\date{Accepted XXX. Received YYY; in original form ZZZ}

\pubyear{2023}

\begin{document}
\label{firstpage}
\pagerange{\pageref{firstpage}--\pageref{lastpage}}
\maketitle

\begin{abstract}
We investigate the transition scale to homogeneity, $R_H$, using as cosmic tracer the spectroscopic sample of blue galaxies from the Sloan Digital Sky Survey (SDSS). 
Considering the spatial distribution of the galaxy sample we compute the two point correlation function $\xi(r)$, the scaled counts in spheres $\mathcal{N}(<r)$, and the fractal dimension $\mathcal{D}_2(r)$ to quantify the homogeneity scale in the Local Universe ($0.04 < z < 0.20$). 
% 
%are independent of cosmological model hypotheses because we calculate the radial comoving distances 
The sample in analysis is compared with {\it random} and {\it mock} catalogues with the same geometry, and the same number of synthetic cosmic objects as the dataset, to calculate the covariance matrix for the errors determination. 
The criteria adopted for the transition-to-homogeneity follows the literature, it is attained when $\mathcal{D}_2(r)$ reaches the $1$ per cent level of the limit value $3$ (i.e., where it reaches $2.97$) as the scale increases. We obtain $R_H = 70.33 \pm 10.74$ Mpc$/h$, at the effective redshift $z_{\text{eff}}=0.128$, 
for a sample containing $150\,302$ SDSS blue galaxies with $0.04 < z < 0.20$. 
Additionally, we perform robustness tests by analysing the homogeneity scale 
in sub-volumes of the original one, obtaining coherent results; we also check for a possible artefact in our procedure examining a homogeneous synthetic dataset as a pseudo-data, verifying that such systematic is absent. 
Because our analyses concentrate in data at low redshifts, $z < 0.20$, we find interesting to use cosmography to calculate the radial comoving distances; therefore in this subject our analyses do not use fiducial cosmological model. 
For completeness, we evaluate the difference of the comoving distances estimation using cosmography and fiducial cosmology.
%This is a model-independent analysis to study the statistical homogeneity in the Local Universe. 
\end{abstract}

%%%%%%%%%%%%%%%%%%%%%%%%%%%%%%%%%%%%%%%%%%%%%%%%%%
%%%%%%%%%%%%%%%%% BODY OF PAPER %%%%%%%%%%%%%%%%%%
\section{Introduction}\label{introduction}

The Cosmological Principle (CP), at the basis of the concordance model of cosmology, the $\Lambda$CDM model, states that the universe is homogeneous and isotropic in three spatial dimensions on sufficiently large 
scales~\citep{Peacock1999,Bernui08,Scrimgeour12,Kashino2012,Rath2013,%
Polastri2017,Tarnopolski2017,Aluri2017,Marques2018,Chiocchetta2021,Goyal2021,Khan2022,Khan2023}. 
However, is still under scrutiny the minimum scale that satisfies the property of homogeneity in the nearby universe~\citep{hogg2005,Scrimgeour12,avila2018}.

Our Local Universe is the part of the universe with the oldest clustered structures. With the gravitational attraction acting during the longest period of cosmic time, $z \simeq 0$, the Local Universe is plenty of large overdense (superclusters)~\citep{Huchra05,tully2019cosmicflows} and large underdense (supervoids)~\citep{Courtois13,Hoffman17} regions, structures that certainly affect the computation of the transition scale to 
homogeneity~\citep{aragon2020scaling,nuza2014cosmic,Hoffman17}. 
In sufficiently large volumes it is expected that the effects of observing clustered matter and voids compensate one to another, and a transition from a non-homogeneous to a homogeneous regime can be found. 
This illustrates the necessity of questioning if the three-dimensional (3D) volume in study is representative of a homogeneous part of the universe, i.e., sufficiently large to contain the homogeneity scale one wants to find. 
We take into account this issue in our analyses.

The study of the statistical homogeneity at different epochs is important to probe the matter clustering evolution of the universe (see, e.g.,~\cite{Marques2020,Garcia2021}). 
Large-scale galaxy redshift surveys offers a unique opportunity to probe the 3D homogeneity, and several analyses of homogeneity with diverse cosmic tracers are being carried out~\citep{york2000sloan,Scrimgeour12}. 
\cite{hogg2005} used a sample of luminous red galaxies with redshift 
$0.20 < z < 0.35$ to obtain a homogeneity scale of $\sim \!\!70$ Mpc$/h$, using a $\Lambda$CDM fiducial cosmology to determine 
3D distances to perform the analysis. 
\cite{Scrimgeour12} used the WiggleZ survey, a spectroscopic survey of over $200\,000$ blue galaxies in a cosmic volume of $\sim \!\!1\ \text{Gpc}^3/h^{3}$; they analysed the homogeneity scale in the redshift 
interval $0.1 < z < 0.9$ obtaining $\sim \!\!70$ Mpc$/h$. \cite{ntelis2017exploring} studied the homogeneity scale using the CMASS galaxy 
sample of BOSS spectroscopic survey in a redshift range of $0.43 < z < 0.7$ and 
obtained a homogeneity scale around $60$ Mpc$/h$. 
The analyses of \cite{laurent201614} of the BOSS quasars sample, with redshifts 
$2.2 < z < 2.8$, conclude that the matter distribution is compatible with 
homogeneity up to $z = 2.8$ (see also~\cite{Goncalves2021}). 
Notice that these results were obtained assuming a fiducial cosmological model to calculate the 3D distances. 
Two-dimensional (2D) analyses, using astronomical data projected on the celestial sphere, can also be done to find the 2D angular scale of 
homogeneity~\citep{gonccalves2018,avila2019}. 
\cite{bharadwaj1999nature} analysed the distribution of galaxies in
three patches in the north galactic hemisphere from Las Campanas Redshift Survey and found $R_H \sim 80$ Mpc$/h$. 
\cite{yadav2005testing} obtained $60 < R_H < 70$ in units Mpc$/h$ studying the two-dimensional strips from SDSS DR1 with $0.08 < z < 0.20$. 
\cite{sarkar2009scale} found $R_H$ between $60$ and $70$ Mpc$/h$ with $0.04 < z < 0.11$. 
Other authors, instead, did not find a homogeneity scale (see, 
e.g.,~\cite{joyce1999fractal,labini2011very,park2017cosmological}).

%\newpage
This work studies the homogeneity scale of a sample of SDSS blue galaxies in the Local Universe $0.04 < z < 0.20$.
%\footnote{we perform analyses for $z > 0.04$ to avoid large peculiar velocities at $z \approx 0$, a systematic effect that can bias our homogeneity scale measurement} 
Our analyses comprise some features that make it different from previous ones: 
(i)~our approach does not assume a fiducial cosmology to calculate the comoving radial distances of the galaxies in analysis (in units Mpc$/h$); instead we use cosmography to determine these quantities 
(however, for the sake of completeness, we evaluate the difference between both distances estimations, i.e., with cosmography and using the concordance cosmological model $\Lambda$CDM); 
(ii) we employ a novel cosmic tracer, and this is important to validate analyses reported in the literature done using other 
tracers~\citep{hogg2005,Sarkar2009,Calcina2018}; 
(iii) we perform substantial robustness tests: to check for possible bias in our procedures with a {\it pseudo-data} test; for error analyses {\it via} covariance matrix evaluation, to check the randomness of our random data set~\citep{Scrimgeour12,ntelis2017exploring,deCarvalho20,deCarvalho21}; additionally, for consistency, we analyse the homogeneity scale in sub-samples of the original one.

Importantly, in many approaches 
the hypotheses regarding local homogeneity and local isotropy are often implicitly assumed; 
here, before to proceed with our analyses we detail explicitly where they are 
used: \\
(a) cosmography is based on the hypothesis that cosmic objects obeys local isotropy, therefore independent on their position on the sky, all 
objects have their radial comoving distances given by equation (\ref{HLlaw}), 
and obeys local homogeneity in the sense that independent of their 3D spatial position with redshift value $z$, again, all objects have their radial comoving distances given by equation (\ref{HLlaw})\footnote{Cosmography certainly assumes that the speed of light $c$ is a universal constant, and $H_0$ is the Hubble constant measured by any observer at $z=0$.}; \\
(b) the Feldman, Kaiser, and Peacock correction, to be described below, also assumes local homogeneity and isotropy as cosmography does; \\
(c) the mock catalogues use fiducial cosmology to be produced, i.e., $\Lambda$CDM, but their use is restricted to the estimation of the statistical uncertainties, through the covariance matrix, in the homogeneity scale measurement.

According to these items, our analyses depend on the hypotheses underlying cosmography, in this sense, they are model-dependent on this theory. 
%However, the radial distances for analyses are not calculated using fiducial cosmology.

%Our study focuses on a test of the CP, that states the universe is statistically homogeneous and isotropic in 3-dimensions (3D), at sufficiently large scales (how large? this is the aim of our analyses). 
The analyses that we perform are suitable for tests of the CP, more specifically, for the cosmic homogeneity at large scales. 
However, it is important to note that these measurements could be suitable probes for studies of alternative scenarios like modified gravity models~\citep{Alam21,Bernui23,Ribeiro23}, the horndessence model~\citep{Linder21}, or extensions of general relativity and $\Lambda$CDM model, such as action theories~\citep{Ntelis23}.

In section \ref{Astronomical data} we present our data and the cuts we did 
in the sample before to apply our methodology; 
in section \ref{Methodology} we explain how to measure the scale of homogeneity given a sample of a cosmic tracer. 
In section \ref{Results and Discussions} we present our results and in section \ref{Conclusions} our conclusions.

%%%%%%%%%%%%%%%%%%%%%%%%%%%%%%%%%%%%%%%%%%%
\section{Astronomical data}\label{Astronomical data}
\label{data}

The Sloan Digital Sky Survey (SDSS) is an international collaboration that performed precise, large, and deep mapping of the universe.
This astronomical survey implemented spectroscopic and imaging observations over large areas of the sky with a $2.5$ m telescope equipped with a large-format mosaic CCD camera to image the sky in five optical bands, $u, g, r, i, z$, and two digital spectrographs~\citep{york2000sloan}.

%more than one fourth of the sky 
The total area covered is about $10\,400$ deg$^2$~\citep{ross2015clustering, alam2015eleventh}\footnote{\url{https://www.sdss4.org/dr12/scope/}} 
and the area covered in the Northern Galactic Hemisphere (NGH), from where we shall select 
our sample for analyses, is around $\sim \!7\,000$ deg$^2$. 

%During the observational campaigns diverse problems, generically termed as systematic effects, arose. They were accounted for in the form of weights applied to the data for selected targets: i) Due to fiber diameter, it's not possible to assign optical fibers on the same plate to two targets closer than 62'', so it is needed to weight this with the close-pair weight, $w_{cp}$ for this effect, ii) targets for which the redshift measurement failed are identified and there's the $w_{noz}$ weight for that, iii) there's a dependence of the observed galaxy number density with the stellar surface density, especially in the equatorial plane, so there's $w_{star}$ to account for that, iv) the observed local galaxy density is also correlated with the {\it seeing} (atmospheric effect), so there's $w_{see}$ for this effect and finally, v) in order to reduce the variance of the two-point correlation function estimator, we use the FKP weight, $w_{FKP}$ that we will explain in more detail in the next section. For more details of the weighting scheme see \cite{ntelis2017probing, reid2016sdss}.

The initial sample is part of the twelfth data release (DR12) from SDSS and consists of $289\,440$ blue star forming galaxies selected from the colour-colour diagram using the
$u, g$, and $r$ broad-bands. 
The galaxies were selected following the criterion $0.0 < g-r < 0.6$ and $0.0 < u - r < 2.0$, in the diagram $g-r\ \times\ u-r$. The magnitudes of each galaxy were corrected for galaxy extinction and for intrinsic reddening (for details about the sample selection and magnitude correction see~\cite{avila2019,deCarvalho21}).

\begin{figure}
    \centering
    \includegraphics[scale=0.4]{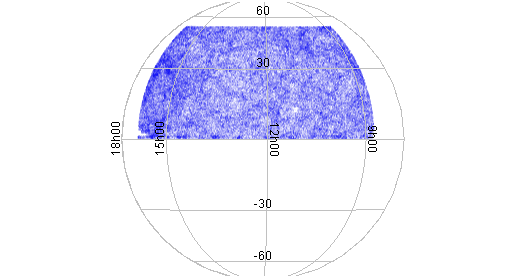}
    \caption{Projection of the sample selected for analyses, $150\,302$ blue galaxies from SDSS, onto the celestial sphere.}
    \label{fig:footprint}
\end{figure}

\begin{figure}
    \centering
    \includegraphics[scale=0.4]{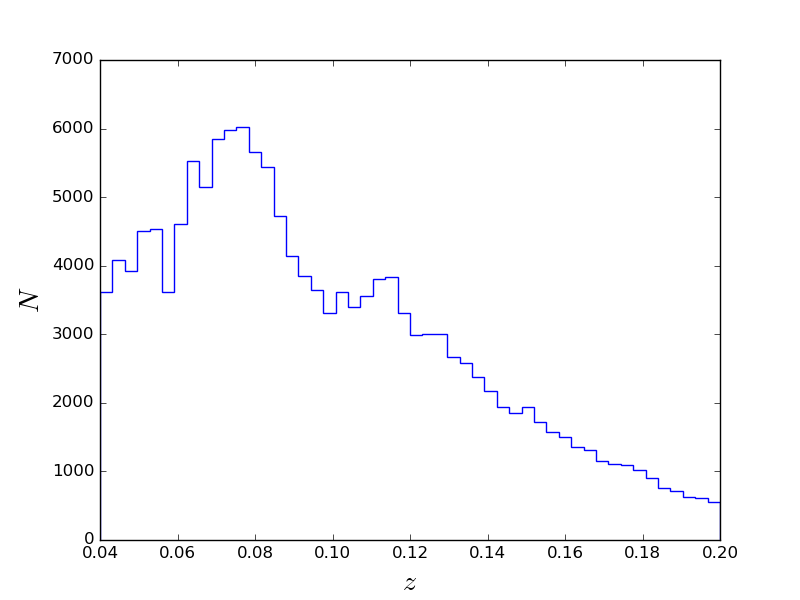}
    \caption{Histogram of the sample in analysis, containing $150\,302$ blue galaxies from SDSS in the redshift interval $0.04 < z < 0.20$.}
    \label{fig:histograma}
\end{figure}

The final sample, used in our analyses, whose footprint is plotted in 
figure \ref{fig:footprint}, has $150\,302$ blue galaxies and was obtained after cuts in angular coordinates and redshift. 
The cut in angular coordinates was applied in order to keep us only with the surveyed region in the NGH, to discard disconnected patches and trim the sample boundaries because the method is sensitive to irregularities at the edges; 
after this, the region for analysis is RA $= [128.91, 243.51]$ deg and DEC 
$= [0.00, 52.71]$ deg. 
Then, the final region after these cuts is $\sim 6\,000$ deg$^2$. 
The cut in redshift leaves the sample for analysis with redshifts in the interval $0.04 < z < 0.20$. 
The redshift selection obeys two reasons: 
i) to avoid large overdense and underdense structures present at low redshifts, effect that could bias our analyses, and ii) in order to apply cosmography to calculate radial comoving ditances, in units of Mpc$/h$; to ensure the validity of our approximation we restrict the analyses for galaxies with $z < 0.20$ 
(see figure~\ref{fig:cosmo_diff}). 
Besides this, for large redshifts the number of observed objects falls rapidly producing a decrease in the number density. 
%, and our aim is to look for a number density for robustness. 
The sample volume is approximately $0.1 \,
\text{Gpc}^{3}/h^{3}$. 
The histogram of the sample data to be analysed, $150\,302$ blue galaxies, 
can be seen in the figure~\ref{fig:histograma}. 

The effective redshift of the sample is $z_{\text{eff}} = 0.128$ and was calculated via \cite{carter2018low}:
\begin{equation}
z_{\text{eff}} = \dfrac{\sum_{i=0}^{Ng} z_i w_i}{\sum_{i=0}^{Ng} w_i}.
\end{equation}

where the weights $w_i$ are defined in equation~(\ref{eq:10}).

\subsection{{\it Randoms}}
The {\it randoms} were created with the code \texttt{randomsdss 0.5.0}\footnote{\url{https://github.com/mchalela/RandomSDSS}} and are used to probe homogeneity comparing them with the data. The {\it randoms} have the same shape, volume and number of points of the data (the {\it randoms} also account for the effect that for larger redshifts we have lower densities by experimental limitation).

\subsection{{\it Mocks}}\label{mocks}
In this work we adopt log-normal simulations to calculate the covariance matrix for the $\mathcal{D}_{2}(r)$ estimator. We used the public code described in \citep{agrawal2017generating}\footnote{\url{https://bitbucket.org/komatsu5147/lognormal_galaxies}}. The code generates mock catalogues in redshift space assuming a log-normal probability density function of galaxy and matter density fields.

In table~\ref{table1} we show all the input parameters needed to generate our log-normal simulated catalogues. In the first column we present the survey configuration: the box dimensions, $L_{x}$, $L_{y}$, and $L_{z}$, the number of galaxies, $N_g$, the redshift at which we generate the input power spectrum, $z$, and the bias, $b$. The matter power spectrum, $P(k)$, is calculated using Eisenstein \& Hu (EH) transfer function~\citep{Eisenstein98}. 
The code uses this approach by default, in case one does not provide a table of $P(k)$ 
values are calculated externally (see \cite{Avila21} for a check of the accuracy of the EH approach).
All these parameters were chosen in order to reproduce our blue galaxies sample. 
In the second column of table~\ref{table1} we observe the cosmological parameters, 
as given by the~\cite{Planck18}.

After generating the mocks, we convert their cartesian coordinates into spherical coordinates, we choose the same footprint and the same redshift range as the sample in analysis, and finally we randomly select $150\,302$ cosmic
objects (to match our data catalogue). 

The covariance matrix for the $\mathcal{D}_2(r)$ estimator, that we define in the next section, is given by 
\begin{equation}
C_{ij} \equiv \dfrac{1}{K-1}\sum^K_{k=1}\left(\mathcal{D}_2^k(r_i)-\overline{\mathcal{D}_2}(r_i)\right)\left(\mathcal{D}_2^k(r_j)-\overline{\mathcal{D}_2}(r_j)\right) \,,
\end{equation}

\noindent where $K$ stands for the number of mocks, $\mathcal{D}_2^k(r_i)$ 
is the $i^{th}$ bin of the $k^{th}$ mock's correlation dimension, and  $\overline{\mathcal{D}_2}(r_i)$ is the $i^{th}$ mock's correlation dimension averaged in the bin. 
The square root of the elements on the main diagonal give us the error bars 
for the plot $\mathcal{D}_2(r)$ versus $r$.

\begin{table}
\caption{Survey configuration and cosmological parameters used to generate the set of $N_{\mbox{s}} =100$ log-normal realisations, also termed {\it mock} catalogues.}
\centering
%	\begin{tabular}{| l | l |} <- original de Felipe
\setlength{\extrarowheight}{0.2cm}
	\begin{tabular}{c|c}
	\hline
	Survey configuration          & Cosmological parameters         \\ 
\hline
	$z=0.128$                              & $\Omega_{c}h^{2}= 0.1202$       \\  
%		\hline
	$b=1.0$                              & $\Sigma m_{\nu}=0.06$         \\
%		\hline
	$N_{g}=10\,000\,000$     & $n_{s}=0.9649$                         \\
%		\hline
	$L_{x}=1\,800$                     & $\ln(10 A_{s})=3.045$                  \\
%		\hline
	$L_{y}=1\,800$                     & $\Omega_{b}h^{2}=0.02236$       \\
%		\hline
	$L_{z}=1\,800$                     & $h=0.6727$       \\ \hline                          
	\end{tabular}
\label{table1}
\end{table}

%%--------------------------------------------------
\section{Methodology}\label{Methodology}

\subsection{Correlation Dimension $\mathcal{D}_2(r)$}

Spatial homogeneity is understood in a statistical sense, as the minimum length scale beyond which any 3D region of volume $V$ contains basically the same number of cosmic objects (this concept is analogous to what is done in fluids, see e.g.~\cite{stoeger1987relationship}).

Since the pioneering work of~\cite{Scrimgeour12}, the most common method to measure the transition length scale to homogeneity is the measurement of the fractal dimension $\mathcal{D}_2(r)$ (also called fractal correlation dimension, or simply {\em correlation dimension}). 
This is a statistical tool to quantify the transition to homogeneity
on a data sample. 

In a statistically homogeneous distribution of objects (uniform distribution) we expect a number of them (galaxies or more generally, particles) proportional to the volume and, consequently, proportional to $r^3$ , where $r$ may be thought as a radius of a sphere~\citep{ntelis2017exploring}
\begin{equation}\label{eq:1}
N(<r)\ \propto\ r^3.
\end{equation}
In a more general situation, a distribution may be, or may be not, homogeneous (even fractal). 
In this case we expect a number of objects proportional to $r^{D_2}$
\begin{equation}\label{eq:2}
N(<r) \propto r^{D_2},
\end{equation}
where $D_2$ is the so called fractal dimension, where 3 represents homogeneity for a three dimensional distribution. Fractions means fractal distributions.
The fractal dimension $D_2$ is obtained from
\begin{equation}\label{eq:3}
D_2(r) \equiv \frac{d\ ln\ N(<r)}{d\ ln\ r} \,.
\end{equation}
${D}_2(r)$ quantifies how the counting number varies with the scale. Applying the analysis just discussed directly presents a challenge due to the inherent limitation in the volume of the sample, regardless of its size. This challenge arises because as the counting spheres increase in size, they progressively encompass larger regions beyond the sample, mistakenly counting these areas as voids. Consequently, this introduces bias into our analysis. The solution to solve this problem is to compare our sample of data (data sample) with an artificial sample ({\it random} sample) with a uniform random distribution (the most "homogeneous" distribution possible for a discrete distribution) with the same number of points, same shape and volume. Thus, we modify our estimators and use the following expressions~\citep{Scrimgeour12,laurent201614, ntelis2017exploring}

\begin{equation}\label{eq:4}
\mathcal{N}(<r) \equiv \frac{N(<r)}{N_R(<r)} \,,
\end{equation}
\begin{equation}\label{eq:5}
\mathcal{D}_2(r) \equiv \frac{d\ ln\ \mathcal{N}(<r)}{d\ ln\ r}+3 \,,
\end{equation}
where $N(< r)$ and $N_R(< r)$ are the number of galaxies counted inside a sphere 
of radius $r$ in the data catalogue and in the random sample, respectively. 
During a counting round, we tally the number of galaxies within a sphere of 
radius $r$, with a chosen galaxy serving as the centre of the sphere. 
Then, we consider the next sphere centred at the next galaxy, and so on, in this 
way the sphere's centre always coincides with the position of each galaxy in the 
catalogue. 
We sum the number for the round (for all galaxies as centre for a given $r$) and 
the resulting number is $N$. $\mathcal{N}(<r)$ is the so called scaled counts in 
spheres. 
Clearly, for homogeneous distributions $\mathcal{N}=1$ and $\mathcal{D}_2=3$. 

To compute the correlation function, it is essential to convert redshift measurements into distances, and then calculate the comoving separation between any pair of cosmic objects in the redshift space~\citep{Sanchez11}
\begin{equation}
    r(z_{1}, z_{2}, \theta) = \sqrt{\chi(z_{1})^{2} + \chi(z_{2})^{2} - 2\chi(z_{1})\chi(z_{2})\cos\theta},
\end{equation}
where $\chi(z)$ is the radial comoving distance at redshift $z$, and $\cos\theta$ is the angular distance between the pair of cosmic objects, with redshifts $z_{1}$ and $z_{2}$. 
Because a significant portion of our sample lies within the Local Universe, we employ the Hubble-Lema\^{\i}tre relation to calculate the radial comoving distance to us of the object at redshift $z$~\citep{Kaiser87,Erdogdu06}
\begin{equation}
\chi(z) = \frac{c z}{H_{0}} 
+ \frac{\langle V_p(\chi) \rangle}{H_{0}},
\end{equation}
where $\langle V_p \rangle$ represents the peculiar velocity field of the object in analysis, that is 
%we parameterise the Hubble constant as $H_{0}\equiv 100 \,h$ km/s/Mpc. 
the local motion of the cosmic object due to the gravitational field of its surrounding matter distribution~\citep{Kaiser87}; on average peculiar velocities reach 
$\langle V_p \rangle \simeq 600$ km/s~\citep{Hoffman17}.
Since the redshifts of our sample satisfies $z > 0.04$, then 
$cz \gg \langle V_p \rangle$ and the approximate equation 
%which means that the recession velocities are larger than $12 000$ km/s, the peculiar velocities field is a sub-dominant effect and we can make the following approximation
\begin{equation}\label{HLlaw}
\chi(z) \simeq \frac{c z}{H_{0}} \,,
\end{equation}
can be considered; this relation implies that redshift and real spaces are close to each other, particularly for scales close to the homogeneity scale. 
%As we are interest in the large scales, where the transition to homogeneity appears, we do not expect significant influence of peculiar velocities in our results. 

\begin{figure}
\centering
\includegraphics[scale=0.6]{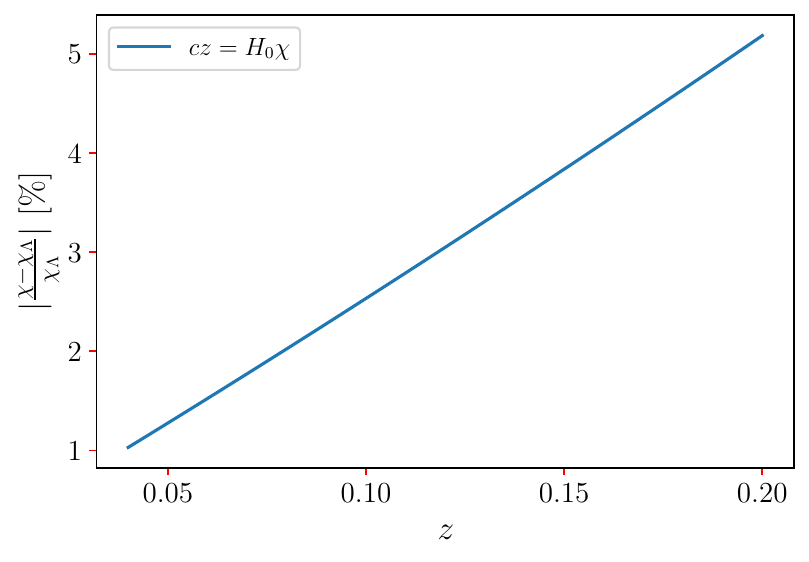}
\caption{Analysis of the relative difference, in percentage, for comoving distance calculations between the cosmographic equation~(\ref{HLlaw}) and the $\Lambda$CDM model. As observed, the maximum relative difference is around $5\%$.}
    \label{fig:cosmo_diff}
\end{figure}

In addition, equation~(\ref{HLlaw}) is a good approximation for low redshift objects $z < 0.2$; in figure \ref{fig:cosmo_diff} we show the relative difference, in percentage, between our cosmography equation and the $\Lambda$CDM model. 
For our sample with $z<0.2$, the maximum relative difference is around $5\%$. 
This relation is useful as the model assumed to obtain $\mathcal{D}_{2}(r)$ is derived from linear perturbation theory without peculiar velocity corrections~\citep{Peebles80}. 
More details can be seen in section~\ref{LCDM model}.

%%----------------------------------------------
\subsection{The 2-point correlation function}

It is possible to show that we can get the same information obtained with the counting spheres method from the set of paired distances between galaxies in the data and the set of paired distances between points in the {\it random}. 
Nevertheless, it's possible to show that there's a function which depends only on these paired distances which is able to characterise spatially a sample in terms of it's numerical density. The function is the so called 2-point correlation function (2PCF) $\xi(r)$. 

The most intuitive definition of the 2-point correlation function (2PCF) is given by the Peebles-Hauser (PH)
estimator~\citep{PH74}
\begin{equation}\label{eq:6}
\xi(r) = \frac{DD(r) - RR(r)}{RR(r)} \,,
\end{equation}
where $DD(r)$ is the number of pairs whose comoving distance between them is in the bin $r$ and $r+\Delta r$ for the {\it data} sample, and $RR(r)$ has the same meaning for the {\it random} sample.

A more suitable 2PCF estimator was proposed 
by~\cite{landy1993bias} (LS), the estimator that we use in our analyses, is given by 
\begin{equation}\label{eq:7}
\xi(r) = \frac{DD(r) - 2DR(r) + RR(r)}{RR(r)}\,,
\end{equation}
\noindent{where} $DR(r)$ represents the number of pairs for which the distance between them falls within the range of $r$ to $r+\Delta r$, considering one galaxy from the data sample and one point from the random sample.
The terms $DD$ and $RR$ have the same meaning that in earlier definition. 
In the Appendix~\ref{apendiceA1} we perform, for comparison, analysis of the homogeneity scale 
using the PH estimator. 
%\textcolor{violet}{Se eu revisei corretamente, o apendice A2 é citado primeiro que o A1. Bruno, por favor, verifique a ordem de citação dos apendices no texto.}

$DD$, $RR$, and $DR$ are normalised by the number of pairs. 
$DD$ and $RR$ are divided by $n(n-1)/2$ and $DR$ is divided by $n^2$, where $n$ is the number of galaxies in the data or in the {\it random} catalogues. 
As mentioned above, we are considering the same number of cosmic objects in the data set, and in the random and mock sets. 

The measurements of $\xi(r)$ were obtained using the code TreeCorr\footnote{\url{https://github.com/rmjarvis/TreeCorr}} by \cite{jarvis2004skewness} and were accounted for the 
Feldman, Kaiser, and Peacock (FKP) correction as described 
in~\cite{feldman1993power} 
(see, e.g.,~\cite{deCarvalho21} and \cite{reid2016sdss}) . 
The $\xi(r)$ estimator accounting for FKP correction is given by 
\begin{equation}\label{eq:8}
\xi(r) = 1+\frac{DD(r)}{RR(r)}\left(\frac{n_R}{n_D}\right)^2-2\frac{DR(r)}{RR(r)}\left(\frac{n_R}{n_D}\right)\,,
\end{equation}
\noindent where the ratio $n_R/n_D$ is given by
\begin{equation}\label{eq:9}
    \dfrac{n_R}{n_D}=\dfrac{\sum^{N_R}_i w_i}{\sum^{N_D}_j w_j}\,,
\end{equation}
\noindent{and} $w$ is obtained from 
\begin{equation}\label{eq:10}
w_i = \dfrac{1}{1 \,+\, n(z_i)\,P_0} \,,
\end{equation}
\noindent{where} $n(z_i)$ is the number density at $z_i$, and $P_0$ is the power spectrum amplitude. 
In this analysis we use $P_0=10\,000$ Mpc$^3/h^{3}$. 
Therefore, $n_R/n_D$ is a correction that optimises the signal-to-noise ratio (SNR) between the sample variance and the shot noise for the 2PCF, $\xi(r)$, and is obtained from individual weights $w_i$ for each galaxy from the data and random sets, respectively. 

In the context of spatial characterisation of a sample of galaxies, $\xi(r)$ can be interpreted as an excess (or as a missing) probability of finding a galaxy in the neighbourhood of a given position if in this region exists an overdensity (or a void) in relation to a random distribution (uniform distribution). In other words, the probability of finding a galaxy at a location is not independent of whether there is a galaxy in the vicinity of that location or not. This can be expressed mathematically through the following relationship~\citep{peebles1973statistical}
\begin{equation}\label{eq:11}
dP = \overline{n}[1+\xi(\vec{r})]dV \,,
\end{equation}
where $dP$ is the probability of finding a given galaxy in a volume $dV$ located in a position $\vec{r}$ for one distribution with a mean numerical density $\overline{n}$. 
It is possible to simplify the previous expression considering the hypothesis of local isotropy in the Local Universe, the same assumed in the cosmography approach 
and described in the Introduction, section~\ref{introduction}. 
%(see, e.g.,~\cite{Franco2023}). 
In this case the correlation depends only on the distance between the origin of the considered referential and the considered position 
$\xi(\vec{r}) = \xi(r)$, 
\begin{equation}\label{eq:12}
dP = \overline{n}[1+\xi(r)]dV \,.
\end{equation}
One can obtain $\mathcal{N}(<r)$ directly from the 2PCF 
\,$\xi(r)$~\citep{ntelis2017exploring}
\begin{equation}\label{eq:13}
\mathcal{N}(<r) = 1+\frac{3}{r^3}\int_{0}^{r} \xi(r')r'^2 dr' \,.
\end{equation}

\subsection{The $\Lambda$CDM expected function $\mathcal{D}_{2}^{\Lambda\text{CDM}}(r)$: a linear approach}\label{LCDM model}

In our analyses, we will compare our main result, $\mathcal{D}_{2}(r)$, with the concordance cosmological model using the first order approximation, 
termed linear approximation, in the cosmological perturbations. 
First, we define the contrast density, $\delta(\textbf{r}, a)$,
\begin{equation}
\delta(\textbf{r}, a) \equiv \frac{\rho(\textbf{r}, a)-\Bar{\rho}(a)}{\Bar{\rho}(a)},
\end{equation}
where $\rho(\textbf{r}, a)$ is the matter density in the neighbourhood of $\textbf{r}$ at cosmic time $t$ (i.e., when the scale factor is $a(t)$), and $\Bar{\rho}(t)$ is the background matter density at cosmic time $t$.

It is convenient to describe the matter fluctuations in the Fourier space,
\begin{equation}
\tilde{\delta}(\textbf{k}, a) = \int \delta(\textbf{r}, a) \,e^{i\,\textbf{k}.\textbf{r}} \,d^{3}r \,,
\end{equation}
and then one can define the matter power spectrum, $P(\textbf{k},a)$,
\begin{equation}
    P(\textbf{k}_{1}, \textbf{k}_{2}, a_{1}, a_{2}) = \frac{1}{(2\pi)^{3}}\langle \tilde{\delta}(\textbf{k}_{1}, a_{1})\,\tilde{\delta}(\textbf{k}_{2}, a_{2}) \rangle \,.
\end{equation}
Due to the assumed properties of local homogeneity and isotropy of cosmographic theory, 
one can obtain the 2-point matter correlation function from the power spectrum~\citep{Padmanabhan93}
%, $\xi(r,a)$, 
\begin{equation}
    \xi(r, a) = \int P(k, a)\, e^{i\,\textbf{k}.\textbf{r}} \,\frac{d^{3}k}{(2\pi)^{3}} \,.
\end{equation}
In brief, to obtain the correlation dimension $\mathcal{D}_{2}^{\Lambda\text{CDM}}(r, z)$, one has to calculate $\xi(r, z)$, equivalently $\xi(r, a)$, where $a = 1/(1+z)$, from a matter power spectrum model, $P(k, z)$, and use in equation (\ref{eq:13}). 
In this work we produce the theoretical curve for $\xi(r,z_{\text{eff}})$, at $z_{\text{eff}} = 0.128$, using the The Core Cosmology Library 
(CCL)~\citep{chisari2019core}.\footnote{\url{https://github.com/LSSTDESC/CCL}.
}

%\noindent
%\textcolor{blue}{Bruno, Felipe: na linha aqui acima, nao seria apropriado escrever a qual redshift a 2PACF $\xi(r)$ se refere? \\
%Digo isso porque na eq. (21) aparece 
%a dependencia de $\xi$ com z: $\xi(r,z)$.
%}

Generally, a complete template is applied to calculate the correlation dimension, incorporating non-linear effects arising from peculiar velocities that are modelled in the power spectrum~\citep{ntelis2017exploring}. 
However, given that we are dealing with data in the Local Universe, these models still fall short of capturing all the effects that distort the correlation function. Typically, a reconstruction of the density field is performed.

However, because our sample of SDSS blue galaxies satisfies $\delta_{\text{blue gals.}} \simeq \delta_{\text{matter}}$~\citep{Zehavi05,Cresswell09}, 
implying that the bias relative to matter of the SDSS blue galaxies is $b_{\text{blue gals.}} \simeq 1$, 
these cosmic objects can be considered a good cosmological tracer of the underlying matter density for large scales.
%for this one can compare our simple model with the data in analysis for scales larger than 60 Mpc$/h$.

%%--------------------------------------------------------
\subsection{The $1$ per cent Criterium}

Since the work of \cite{Scrimgeour12}, the literature widely adopted the criterium of $1$ per cent method for determining a
homogeneity scale $R_H$ where the transition to homogeneity occurs. This method has the advantage of being easier to compare with theory and between surveys. 
The method consists on fitting a model-independent curve to the data, and find the value of $r$ at which this intercepts a horizontal line at $1$ per cent ($2.97$) from $\mathcal{D}_2(r) = 3$. 

\subsection{Obtaining the transition scale to homogeneity}\label{param_model}

A well-defined methodology needs to be established to determine, as accurate as possible, where the transition to homogeneity in the $1$ per cent criterium occurs. 
To find this transition to homogeneity scale some trial functions, like splines, are considered to fit the data; the difference among the several approaches observed in the literature is that some analyses consider the fit of the data close to the transition scale~\citep{Scrimgeour12}, while others fit a large portion of the data set~\citep{avila2018}, both for $\mathcal{N}(<r)$ and for $\mathcal{D}_2(r)$. 
In general, parametric and non-parametric approaches works fine to solve this problem.

Because our $\mathcal{D}_2(r)$ data exhibit a smooth transition to homogeneity, a simple analytic curve can determine the value of $R_{H}$ and its respective error. The only condition one imposes is that this function has as limit the value $3$, which corresponds to the asymptotic limit of $\mathcal{D}_2(r)$. 
Thus, we have
\begin{equation}\label{def-f}
f(x) \equiv A\left[1-\exp{\left(\frac{B}{x}\right)}\right] + 3 \,,
\end{equation}
where $A$ and $B$ are constants to be determined.

The benefit of adopting an analytic function is the convenience of propagating the error and getting roots of the function. Once the function parameters are determined using a $\chi^2$ minimisation, the error of the function $f(x)$ is, approximately 
\begin{equation}\label{eq:15}
   \sigma_{f(x)} \simeq \sqrt{\left\{\left[1-\exp{\left(\frac{B}{x}\right)}\right]~\sigma_{A}\right\}^{2} +  \left[-\frac{1}{x}A\exp{\left(\frac{B}{x}\right)}~\sigma_{B}\right]^{2}}.
\end{equation}
As for the calculation of $R_{H}$, we have the following relation
\begin{equation}\label{eq:16}
A\left[1-\exp{\left(\frac{B}{R_{H}}\right)}\right] + 0.03 = 0 \,,
\end{equation}
where we use the equation (\ref{def-f}) in the $R_{H}$ point where $f(R_{H})=2.97$. 
Thus, isolating $R_{H}$ we have
\begin{equation}\label{eq:17}
R_{H} = \frac{B}{\ln{\left(1+\frac{0.03}{A}\right)}} \,,
\end{equation}
where the error is, approximately 
\begin{equation}\label{eq:18}
\sigma_{R_{H}} \simeq \sqrt{\left[\frac{R_{H}^2}{B}\frac{3}{A(100A+3)}~\sigma_{A}\right]^{2} + \left(\frac{R_{H}}{B}~\sigma_{B}\right)^{2}} \,.
\end{equation}

Note that our analytic function resembles the definition of $\mathcal{D}_{2}(r)$, equation (\ref{eq:5}), where the logarithm derivative of $\mathcal{N}(<r)$ is approximate to a exponential function. For physical reasons, we do not expect that our function fit small scales, where non-linear scales dominate and the correlation function is well described by a power-law~($\mathcal{D}_{2}(r)$ is related to $\xi(r)$, see equation (\ref{eq:13})). However, as the scales increases, the correlation function can be described approximately by an exponential function~(see equation (27) from \cite{Labini08}). 
Thus we decide to exclude the small scales from our fit, defining a lower limit of $r > 60$ Mpc$/h$.

%%------------------------------------------------

%

%%--------------------------------------------------------------------------
\subsection{Minimum 3D volume for homogeneity analyses}\label{secao3.6}

One of the important questions about large-scale analyses intended to extract cosmological information from a 3D distribution of cosmic objects is to verify if the 3D volume in study is large enough to obtain a result that is significantly robust. 
Regarding the transition from a non-homogeneous spatial distribution to a homogeneous one,~\cite{Coleman92} affirm the following: $R_{S} \gg R_{H}$, that is, the scale containing the cosmological information, e.g. the homogeneity scale, $R_{H}$, must be smaller than the survey scale, $R_{S}$. 

For the estimator used in our analyses, the condition $R_{S}\gg R_{H}$ must be considered because $\mathcal{N}(<r)$ always will reach 1, independently if such condition is satisfied. Consequently, $\mathcal{D}_{2}(<r)$ always will reach 3. 

A methodology was developed to test if the volume analysed contains sufficient information to extract the correct homogeneity scale. 
Following~\cite{laurent201614}, we calculate the pair counts in the random catalogues for a large range in distance. 
From this, we derive and divide by $r^{2}$ and normalise this calculation. 
In this way we obtain the number density of random pairs. 
In the quasar dataset analysis, \cite{laurent201614} considered $10$ per cent as the 
limit for the calculation, that is, when the number density falls below $10$ per cent, 
there is no more sufficiently quasar pairs to obtain a good signal-to-noise for the analysis. As an arbitrary limit,~\cite{ntelis2017exploring} decided to use a $1$ per cent less limit. 
This change was necessary because the sample used in \cite{ntelis2017exploring} 
has a much smaller volume compared with the quasar volume used 
in~\cite{laurent201614}. 

\begin{figure}
\centering
\includegraphics[scale=0.6]{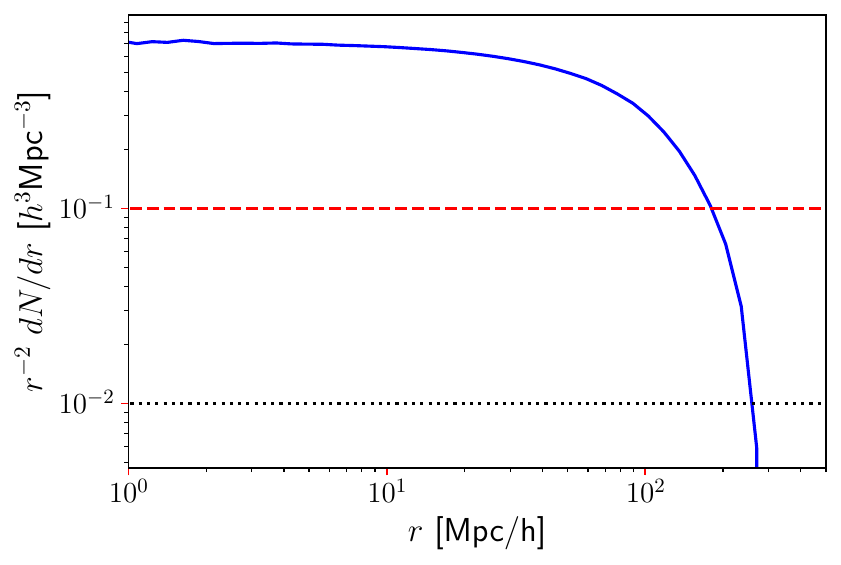}
\caption{Scaled number density of random pairs as a function of the distance between pairs $r$. 
The red dashed line traces the $10$ per cent limit, and the black dotted line, the $1 \%$ limit. See the text for details.}
\label{fig:range}
\end{figure}

In figure \ref{fig:range} we show the result of this calculation. 
The red dashed line traces the $10$ per cent limit, and the black dotted line, the $1$ per cent limit. 
As we are going to show in the next section, the homogeneity scale is reached before the $10$ per cent limit, which indicates the robustness of our result. This is because our sample possible traces much better the dark matter clustering than the intermediate redshift samples like the Luminous Red Galaxies cosmic tracer, therefore, a small homogeneity scale is expected for our sample in study. 

In the Appendix~\ref{apendiceA2} we study, for consistency, the transition to homogeneity scale, $R_H$, in 4 sub-samples of the original data sample. 
This study supports the importance of considering a 3D volume large enough, and containing suitable 
number density of cosmic objects, to obtain a robust measurement of $R_H$.

%%%%%%%%%%%%%%%%%%%%%%%%%%%%%%%%%%%%%%%%%%%%%%%%%%%

\section{Results and Discussions}\label{Results and Discussions}

\begin{figure}
\centering
\includegraphics[scale=0.6]{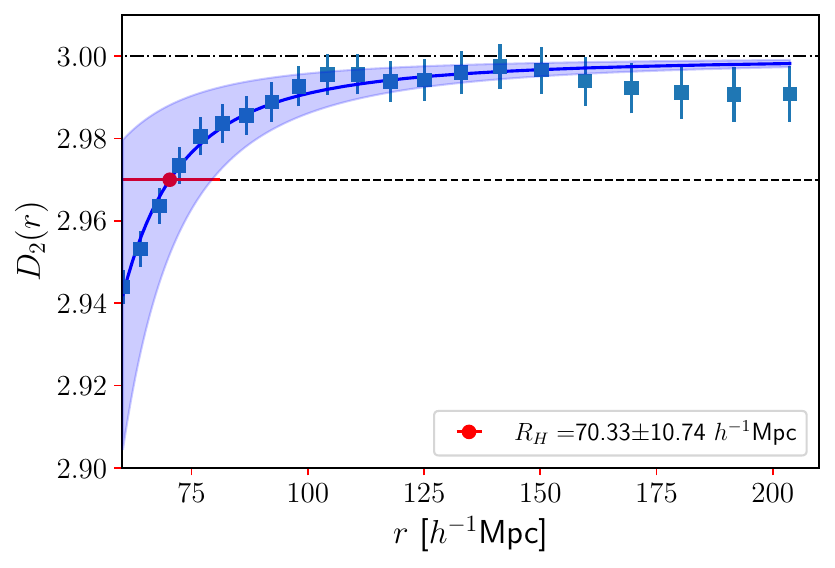}
\caption{$\mathcal{D}_2(r)$ as function of distance between pairs $r$ for the sample with $150\,302$ blue galaxies from SDSS with $0.04 < z < 0.20$. 
The error bars were obtained from the covariance matrix with respect to the $100$ {\it mocks}. The continuous blue line is the best fit according to our parametric model, described in section \ref{param_model}, to obtain the homogeneity scale. The blue shadow region is the 68\% confidence interval obtained from equation~(\ref{eq:15}). 
We show our measurement of the homogeneity scale as a red dot.
}
\label{fig:D2}
\end{figure}

With the well-defined methodology of the previous section, and steps developed there, 
we now discuss the result obtained for $\mathcal{D}_2(r)$ shown in figure~\ref{fig:D2}, 
our measurement of the transition scale of homogeneity. 
We summarise our results in table \ref{tab:best_fit}.

The obtained values of $\xi(r)$ for each bin are mean values referring to 25 randoms.
From the calculated values of $\xi(r)$ we obtain the values of $\mathcal{N}(<r)$ using equation (\ref{eq:13}). The values for $\mathcal{N}(<r)$ again are mean values for the $25$ {\it randoms}.

Finally, from the values of $\mathcal{N}(<r)$ we obtain the values for the fractal dimension $\mathcal{D}_2(r)$, using the equation (\ref{eq:5}) and plot them against the distance between pairs. The error bars were obtained from the bin-bin covariance matrix considering 100 {\it mock} catalogues. The result is shown in figure \ref{fig:D2}.

\begin{table}
\centering
\caption{In this table we summarise our results of the best-fit procedure. We display the 
best-fit parameters found, and the statistical significance of the procedure.}
\begin{tabular}{ll}
Best Fit                             & \hspace{0.8cm} Values                                      \\ \hline
\multicolumn{1}{|l|}{$A$}            & \multicolumn{1}{l|}{$0.0006 \pm 0.0003$}    \\ \hline
\multicolumn{1}{|l|}{$B$}            & \multicolumn{1}{l|}{$272.6391 \pm 28.4099$ Mpc$/h$} \\ \hline
\multicolumn{1}{|l|}{$R_{\text{H}}$} & \multicolumn{1}{l|}{$70.33 \pm 10.74$ Mpc$/h$} \\ \hline
\multicolumn{1}{|l|}{$\chi^{2}$}     & \multicolumn{1}{l|}{6.594}                  \\ \hline
\multicolumn{1}{|l|}{$\nu$}          & \multicolumn{1}{l|}{21 - 2}                 \\ \hline
\multicolumn{1}{|l|}{$\chi^{2}/\nu$} & \multicolumn{1}{l|}{0.347}                  \\ \hline
\label{tab:best_fit}
\end{tabular}
\end{table}

The dashed and dash-dotted black lines represent correlation dimension values of $2.97$ and $3.00$, respectively. The blue square points represent the measured correlation dimension for our sample of blue galaxies, with the associated error bars derived from log-normal simulations. By employing our parametric model, as described in equation (23), and applying the $1\%$ criterion, we determine a value of $R_H = 70.33 \pm 10.74$ Mpc$/h$, displayed as a red dot in figure \ref{fig:D2}. 
The uncertainty in this measurement is computed using the approximation outlined in equation (27).
Our parametric model is represented by the blue curve, fitted in the range $60 < r < 200$ Mpc$/h$. 
The uncertainty, from equation (24), is visualised at a 68\% confidence level (CL) through a shaded blue region. Particularly noticeable is the substantial increase in error as we get deeper into smaller scales. This aligns with our expectations, as our model primarily addresses larger scales. Furthermore, our model effectively captures a smooth transition scale, evidenced by the absence of any discernible discontinuity between the points near 2.97.

%The dash dotted red line indicates the value $2.97$ which differs in $1$ per cent from $3$ indicating the transition to homogeneity as discussed earlier. Adopting this criterion and the method discussed, we obtain a value for the transition scale to homogeneity $R_H = 70.33\pm 10.74$~$h^{-1}$Mpc. 
%A value of $70.40$ $h^{-1}$Mpc was obtained for a flat-$\Lambda$CDM fiducial cosmology using the package CCL \citep{chisari2019core}\footnote{\url{https://github.com/LSSTDESC/CCL}} which can be seen as the dark green line in the figure \ref{fig:D2}.

\begin{figure}
\centering
\includegraphics[scale=0.5]{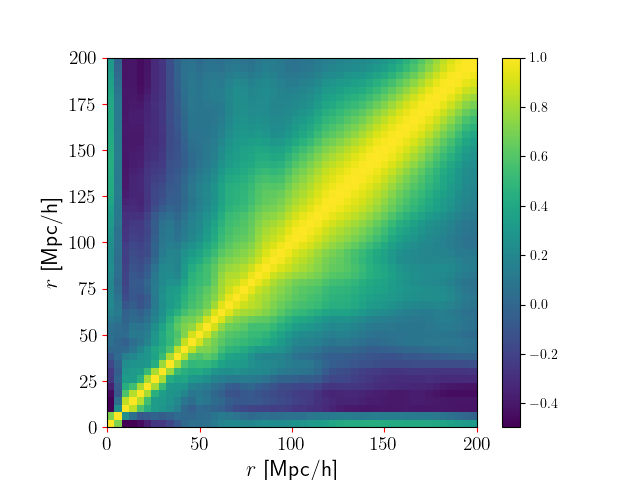}
\caption{Correlation matrix calculated using $100$ mocks to obtain 
the error bars for $\mathcal{D}_2(r)$. 
See the sub-section~\ref{mocks}, and table~\ref{table1}, for details of the generation of the mocks used in this analysis.}
\label{fig:correlation_matrix}
\end{figure}

In figure \ref{fig:correlation_matrix} we present the bin-to-bin reduced covariance matrix, also called correlation matrix, calculated for $100$ {\it mocks} with $50$ bins each in order to obtain the error bars for $\mathcal{D}_2(r)$. 
The correlation matrix is given by 
\begin{equation}
r_{ij}=\dfrac{C_{ij}}{\sqrt{C_{ii}C_{jj}}} \,.
\end{equation}

\begin{figure}
\centering
\includegraphics[scale=0.4]{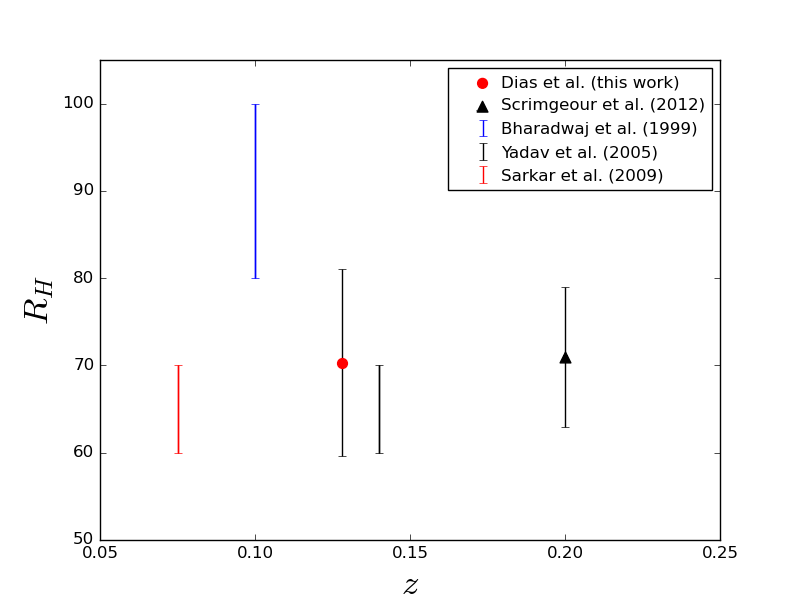}
\caption{Measurements of $R_H$ from diverse analyses at low redshifts: \protect\cite{Scrimgeour12}
obtained $R_H = 71 \pm 8$ Mpc$/h$ for blue galaxies of the WiggleZ survey, with redshift $z \sim 0.20$; \protect\cite{bharadwaj1999nature} found $R_H$ between $80$ and $100$ Mpc$/h$ with $z \simeq 0.1$; \protect\cite{yadav2005testing} studying the two-dimensional strips from SDSS DR1 obtained $60 < R_H < 70$ in units Mpc$/h$ with $0.08 < z < 0.20$;~\protect\cite{sarkar2009scale} found $R_H$ between $60$ and $70$ Mpc$/h$ with $0.04 < z < 0.11$; our measurement of $R_H = 70.33 \pm 10.74$ Mpc$/h$, at $z_{\text{eff}}= 0.128$, is represented as a red dot.}
\label{fig:RHs}
\end{figure}

It is interesting to compare our measurement of the homogeneity scale for the SDSS blue galaxies as a cosmological tracer in the Local Universe, $R_H$ at the effective redshift $z_{\text{eff}}=0.128$, with other measurements reported in the literature. 
This comparison of data at low redshifts is shown in figure \ref{fig:RHs}, where we observe that three of these results are plotted as a range and the other two are plotted as points with its margin of errors. It is important to point out that the measurements in Figure \ref{fig:RHs} were not corrected to obtain the dark matter homogeneity scale, which decreases exponentially with redshift. Figure \ref{fig:RHs} is intended to show that our $R_H$ measurement is consistent with other local measurements using other types of dark matter tracers.

\subsection{Comparison of our methodology with 
the fiducial model}

\begin{figure}
    \centering
    \includegraphics[scale=0.55]{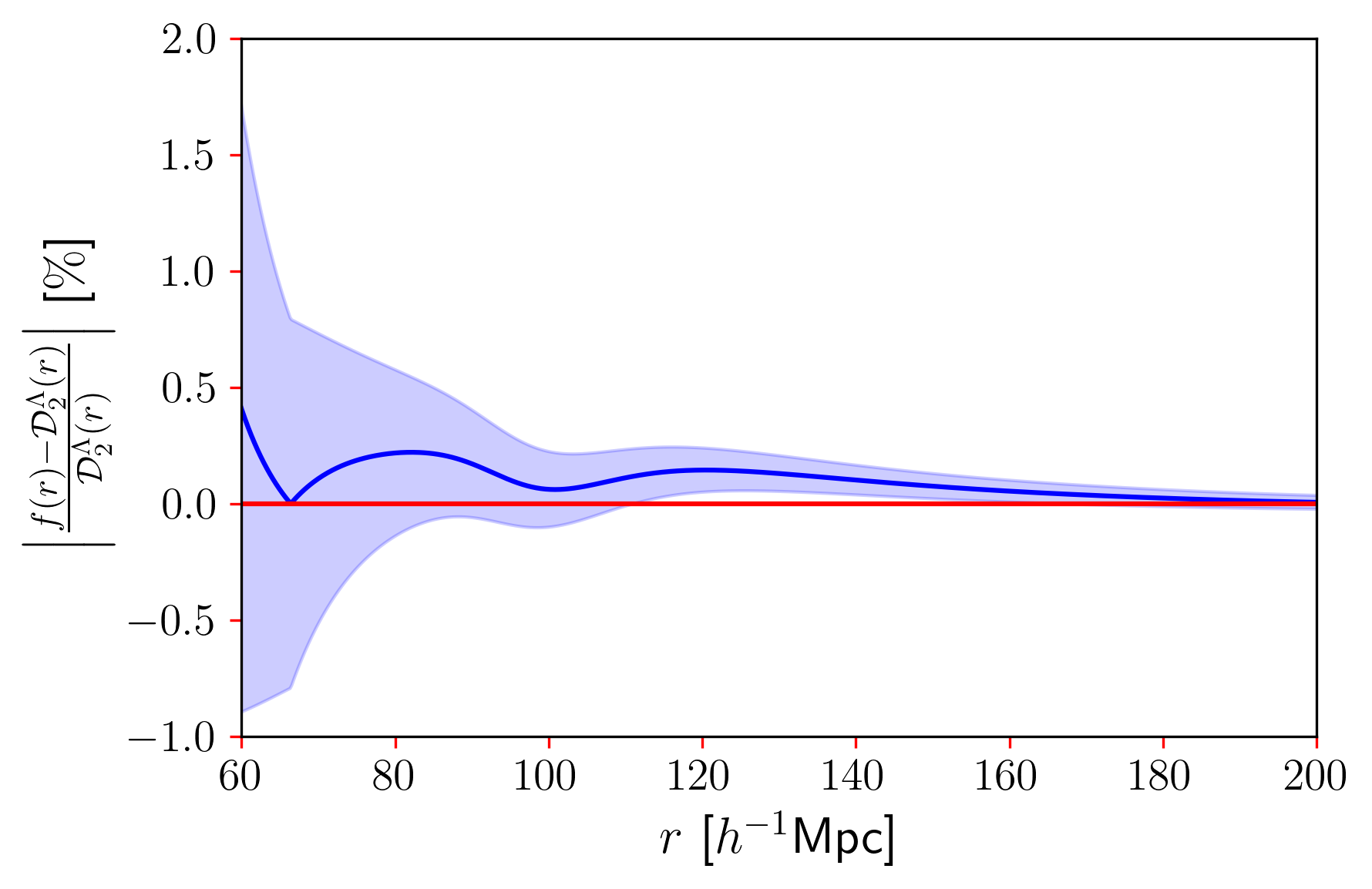}
    \caption{
Analysis of the relative difference between $\mathcal{D}^{\Lambda\text{CDM}}_2(r)$, calculated using the fiducial cosmology $\Lambda$CDM, and our parametric model $f(r)$, given by equation~(\ref{def-f}). 
The shadow region represents the 68\% CL.
}
\label{fig:model_fit}
\end{figure}

In this section, we investigate the parametric model 
we use for fitting the $\mathcal{D}_2$ data, and compare it with the $\Lambda$CDM model in the linear approximation. For this, we calculate the relative difference, in percentage, between our parametric model $f(r)$, given in equation (\ref{def-f}), and the theoretical model for the correlation dimension, $\mathcal{D}_{2}^{\Lambda\text{CDM}}(r)$, derived from equation (22), (18), and (7). In figure \ref{fig:model_fit}, we show this relative difference, our statistical theoretical uncertainty. The shadow blue region is the 68\% CL propagated from the best-fit.

We note from figure \ref{fig:model_fit} a maximum of 0.5\% between the parametric and theoretical model in the best fit range. 
The small difference observed can be explained as follows.
As mentioned in section~\ref{secao3.6}, the SDSS data sample we are analysing provides an excellent cosmic tracer to measure the 
transition scale to homogeneity. 
This is due to the nature of blue galaxies to have a bias relative to dark matter approximately equal to $1$~\citep{Cresswell09,avila2019}. 
In other words, blue galaxies are predominantly found outside high-density regions, unlike other cosmic tracers like Luminous Red Galaxies. 
In addition, a comparison between blue galaxies and extra-galactic sources emitting at 21 cm reveals a minimal difference in the amplitudes of their correlation functions, at the same time these sources have a bias relative to dark matter close to 1~(see, 
e.g.,~\cite{Martin12,Papastergis13,avila2018}). 

\begin{figure}
\centering
\includegraphics[scale=0.55]{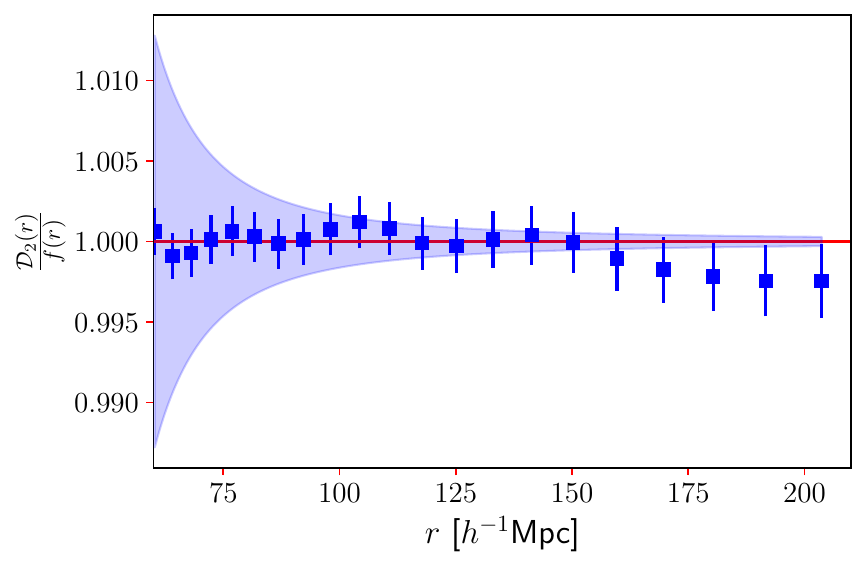}
\caption{Analysis of the fraction between our measurement $\mathcal{D}_{2}(r)$ and the parametric model $f(r)$ with the best fit parameters. 
The 68\% CL shadow region is the fraction between the uncertainties in these quantities.
For scales larger than 100 Mpc$/h$ the data uncertainties dominates. 
For scales smaller than 100 Mpc$/h$ the theoretical 
approach dominates, a consequence of the fact that the parametric model is not 
suitable for small scales.}
\label{fig:error_fit}
\end{figure}

Besides the theoretical uncertainties related to the $\Lambda$CDM model, we also compare our parametric model with uncertainties in the data. 
This comparison is shown in figure \ref{fig:error_fit}. 
This figure illustrates how the parametric model --with the best-fit parameters-- behaves with respect to our measurements of 
$\mathcal{D}_{2}(r)$. 
The shaded region at the 68\% CL represents how uncertain our parametric model is when compared to the theoretical model. 
In general, the theoretical model presents smaller uncertainties compared to the observational data.

In the figure, we can see something interesting: for scales larger than 100 Mpc$/h$, the data uncertainties are larger than what the theoretical model suggests. 
On the other hand, for smaller scales, the theoretical uncertainties are larger. 
This makes sense because our parametric model was designed for large scales. 
The figure also helps to explain why the reduced $\chi^{2}$ value is low, as seen in table \ref{tab:best_fit}. 
Until $\sim \!150$ Mpc$/h$, the data points 
in figure \ref{fig:error_fit} are consistently below the value $1$. 
This means that the function $\mathcal{D}_{2}$ does not get close to the value $3$ that our parametric model predicts because the 3D volume occupied by the sample under analysis 
is not large enough, and does not contain enough number density of cosmic objects, 
for this asymptotic behaviour, $\mathcal{D}_{2} \rightarrow 3$, to occur.

%%%%%%%%%%%%%%%%%%%%%%%%%%%%%%%%%%%%%%%%%%%%%%%%%%%
\section{Conclusions}\label{Conclusions}

Our study focused on a precise test of the CP, which establishes that the universe is statistically homogeneous and isotropic in three spatial dimensions, at large enough scales. 
Our analyses provide a good agreement with the $\Lambda$CDM model, which is based on the CP.

The study of matter clustering, and the effects it produces in the large scale matter distribution, is important to understand the evolution of the universe~\citep{Marques2020,Thiele23,Bargiacchi2023,Dainotti2023,Specogna2023,Staicova2023,Marques23,Avila23}, but it is also related to the (scale of) statistical homogeneity at different epochs (see, e.g.,~\cite{Garcia2021,Avila2022}). 
In this work we explore, analytically, how to quantify the homogeneity scale given the spatial distribution of a catalogue of SDSS blue galaxies. 
The criteria adopted in our analyses for the transition-to-homogeneity follows the literature, it is attained when $\mathcal{D}_2(r)$ reaches the $1$ per cent level of the limit value $3$ (i.e., where it reaches $2.97$) as the scale increases.

Our methodology  is based on cosmographic theory (see, e.g.,~\cite{visser2004jerk}), for this reason our approach assumes the same hypotheses regarding local homogeneity and local isotropy underlying such theory (see section~\ref{introduction} for details~\footnote{
Our analyses depend on the hypotheses underlying cosmography, in this sense, they are model-dependent on this theory.
}). 
%Notice that our analyses depends on the hypotheses underlying cosmography, in this sense, they are model-dependent.
Clearly, these hypotheses do not guarantee the existence of a homogeneity scale, which is why our analyses are interesting. 
We adopted cosmography to calculate the radial comoving distances using equation~(\ref{HLlaw}). Note that this approach is different from diverse analyses reported in the literature that use fiducial cosmology, e.g. $\Lambda$CDM, to calculate such distances. 
For completeness, in figure~\ref{fig:cosmo_diff} we evaluate the difference of the comoving distances estimation using cosmography and fiducial cosmology $\Lambda$CDM.

Summarising, we have investigated the property of statistical homogeneity in the Local Universe by measuring the 3D transition to the homogeneity scale in a sample of \,$150\,302$\, SDSS blue galaxies as a cosmological tracer, with redshifts in the interval $0.04 < z < 0.20$. 
The data sample was compared with random and mock catalogues, with the same footprint and the same number of objects, to compute the covariance matrix for the errors determination. 
From our analyses we conclude that the transition to homogeneity scale, at the effective redshift $z_{\text{eff}}=0.128$, is $R_H = 70.33 \pm 10.74$ Mpc$/h$, a value in full agreement with other measurements reported in the literature, as illustrated in figure~\ref{fig:RHs}.

One interesting issue, that we discuss in the Appendix~\ref{apendiceA1}, is the study of the homogeneity scale using the PH estimator for the calculation of the 2PCF (see figure~\ref{fig:D2PH}), 
analysis that allows us to compare with our result obtained using the LS estimator (see figure~\ref{fig:D2}).

Another important subject is the analyses of the homogeneity scale considering small 3D volumes, that is, sub-samples of the whole data sample investigated. 
In fact, as observed in our analyses in Appendix~\ref{apendiceA2}
the presence of diverse superclusters and large voids in the Local Universe~\citep{Courtois13,tully2019cosmicflows} has an impact in the computation of the homogeneity scale and in its uncertainty measurements, although the results obtained are coherent and statistically compatible within $1 \,\sigma$ CL.

%As a result, medium-size spatial volumes may not give account of the necessary completeness to measure the transition to homogeneity scale. 
%Finally, we have performed robustness tests by analysing the homogeneity scale in sub-volumes of the original one, obtaining coherent results. 

In addition, in the Appendix~\ref{apendiceA3} we check for a possible artefact in our procedure 
that measures the homogeneity scale $R_H$. 
We perform  this test by examining a homogeneous synthetic data, termed pseudo-data, as it would be a data catalogue. 
Our result, displayed in figure~\ref{fig:D2pseudo}, shows that the pseudo-data catalogue is homogeneous, verifying that a possible systematic is absent in our analyses.

In the future, one expects that measurements of the homogeneity scale could be suitable probes for studies and tests for alternative scenarios, like modified gravity models~\citep{Alam21}, horndessence model~\citep{Linder21}, or extensions of general relativity and $\Lambda$CDM, as 
%the existence of actionic fluctuations prediction of 
action theories~\citep{Ntelis23}.

%We obtained a measurement for $R_H$ consistent with other data reported in the literature; we calculated the corresponding values for $\mathcal{N}(<r)$ and $\mathcal{D}_2(r)$ for blue galaxies in the Local Universe using cosmography, which were also reasonable in relation to what was expected by the models and other works in the literature.

%%%%%%%%%%%%%%%%%%%%%%%%%%%%%%%%%%%%%%%%%%%%%%%%%%%
\section*{Acknowledgements}

%the valuable helpful of 
BLD acknowledges his permanent institution, the Instituto Federal de Santa Catarina (IFSC) -- Campus Araranguá, for the leave of absence. 
BLD thanks Agripino Sousa and Camila Nascimento Franco for their help with computing facilities at the Observat\'orio Nacional. 
FA thanks the Conselho Nacional de Desenvolvimento Cient\'{i}fico e Tecnologico (CNPq, National Council for Scientific and Technological Development) and the FAPERJ - Fundação Carlos Chagas Filho de Amparo à Pesquisa do Estado do Rio de Janeiro for their financial support.
AB thanks a fellowship from CNPq. 
%for the grants under which this work was carried out.
We all acknowledge 
the SDSS team for the use of the LSS data survey.

%%%%%%%%%%%%%%%%%%%%%%%%%%%%%%%%%%%%%%%%%%%%%%%%%%
\section*{Data Availability}
 
The data underlying this article will be shared on reasonable request to the corresponding author.


\bibliographystyle{mnras}
\bibliography{example} % if your bibtex file is called example.bib

%%%%%%%%%%%%%%%%%%%%%%%%%%%%%%%%%%%%%%%%%%%%%%%%%%

%%%%%%%%%%%%%%%%% APPENDICES %%%%%%%%%%%%%%%%%%%%%
\appendix

\section{Consistency tests}\label{apendiceA}

Next, we present consistency tests performed to verify the robustness of our results. 
We analyse the homogeneity scale in sub-samples of the original dataset, we use a different estimator for the 2PCF and repeat the analyses displayed in figure~\ref{fig:D2}, and finally, we test the whole analysis procedure considering a {\em pseudo-data} set, i.e., a random homogeneous catalogue as it would be a data catalogue.

%%----------------------------------------------------------------
\subsection{Calculating $\mathcal{D}_2(r)$ with the Peebles-Hauser estimator}\label{apendiceA1}

\begin{figure}
\centering
\includegraphics[scale=0.5]{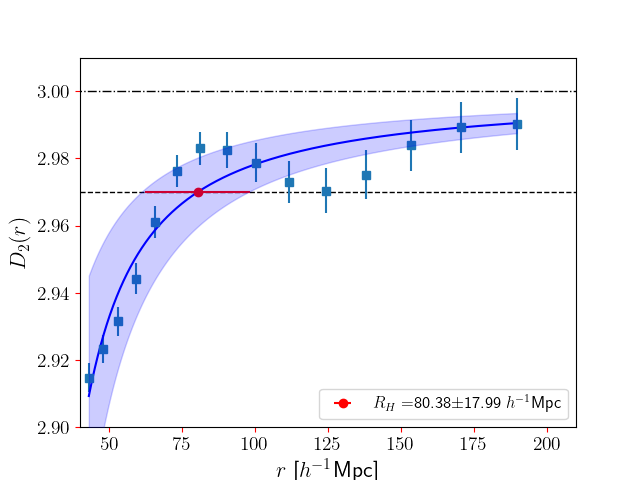}
\caption{Calculation of the scaled counts in spheres, $\mathcal{D}_2(r)$, as a function of distance, $r$, for the original sample with $150\,302$ blue galaxies from SDSS with $0.04 <  z < 0.20$, but this time using the Peebles-Hauser 2PCF estimator. 
The error bars were obtained from the covariance matrix using $100$ {\it mocks}.}
\label{fig:D2PH}
\end{figure}

%\textcolor{red}{Redigido por Armando}\\
Our scope here is to confirm the homogeneity scale measurement using a different estimator, that is, the Peebles-Hauser estimator, sometimes also employed in the literature~\citep{Goncalves2021}. 
One notices in figure~\ref{fig:D2PH} that $\mathcal{D}_2(r)$ shows a deep depression around 120 Mpc$/h$, which corresponds to the feature observed at the same scale in the $\mathcal{D}_2(r)$ of figure~\ref{fig:D2}, which was obtained with the Landy-Szalay estimator of the 2PCF. 

This result confirms that both estimators are statistically compatible, 
although the Landy-Szalay estimator is usually recommended in data analyses 
because it can reach minimal (nearly Poisson) 
variance~\citep{landy1993bias}. 
%recommend an improved estimator (DD - 2DR + RR)/RR, whose variance 
%is nearly Poisson. 
%the Landy-Szalay estimator, which is supposed to reach minimal variance

%%----------------------------------------------------------------
\subsection{Analyses of $R_H$ with sub-samples}\label{apendiceA2}

\begin{figure}
\centering
\includegraphics[scale=0.5]{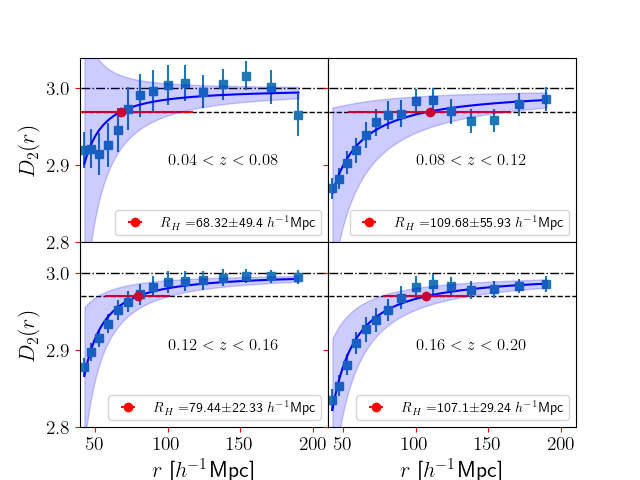}
\caption{Calculation of the transition to homogeneity scale, $R_H$, for various redshift intervals 
corresponding to sub-samples of the original sample investigated 
in this work, and described in section~\ref{data}. 
%The analyses shown in figures~\ref{fig:subsamples1} and~\ref{fig:subsamples2} indicate that our main result displayed in figure~\ref{fig:D2} is robust. 
Considering the uncertainty obtained for these measurements, we conclude that all of them are statistically compatible with our $R_H$ measurement shown in figure~\ref{fig:D2}.
\label{fig:subsamples1}}
\end{figure}

\begin{figure}
\centering
\includegraphics[scale=0.5]{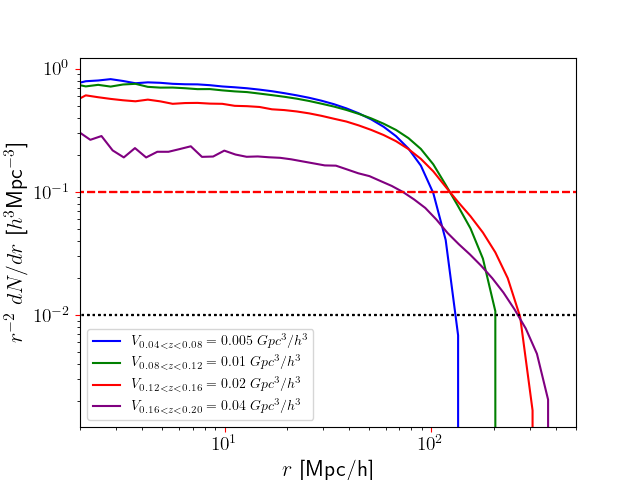}
\caption{
Scaled number density of random pairs as a function of the distance between pairs for the 4 redshift intervals considered in the analyses displayed in figure~\ref{fig:subsamples1}. 
The red dashed line traces the 10 per cent limit, and the black dotted line, the 1 per cent limit. 
For comparison, observe the analysis done in figure~\ref{fig:range} considering the original sample. 
These sub-sample studies complement the analyses shown in figure~\ref{fig:subsamples1}, making evident that data sub-samples may not contain enough 3D volume data to find there, in a robust way, the transition to the homogeneity scale $R_H$.
\label{fig:subsamples2}}
\end{figure}

In this section we inquire what would be the homogeneity scale $R_H$ 
if we apply our methodology to sub-volumes of the original one; 
our results are shown in figures~\ref{fig:subsamples1} 
and~\ref{fig:subsamples2}. 
As observed, the $\mathcal{D}_2(r)$ function calculated in these analyses 
shows an irregular behaviour, suggestive that in these sub-samples 
the presence of galaxy clusters and voids affects considerably the computation of the homogeneity scale $R_H$. 

In figure \ref{fig:subsamples2} we plot the pair density as a function of distance for the 4 redshift intervals considered. For the first interval (0.04 < z < 0.08) we have $60\,293$ points, for the second interval (0.08 < z < 0.12) we have $49\,434$ points, for the third interval (0.12 < z < 0.16) we have $28\,618$ points, and for the last interval ( 0.16 < z < 0.20) we have $11\,957$ points. It is possible to observe that for all the intervals considered we have a characteristic distance, given by the intersection with the dashed line, i.e., the 10 per cent limit in figure~\ref{fig:subsamples2}, corresponding to $\sim 50\%$ 
to that one of the original sample (0.04 < z < 0.20).

%%----------------------------------------------------------------
\subsection{Calculating $\mathcal{D}_2(r)$ with a pseudo-data catalogue: a random set}\label{apendiceA3}

\begin{figure}
\centering
\includegraphics[scale=0.5]{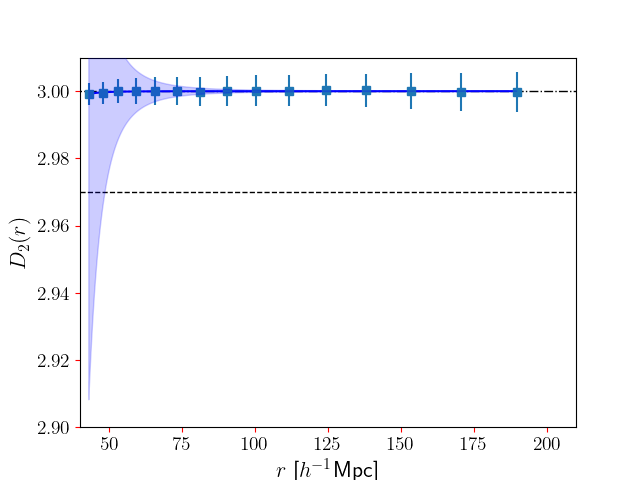}
\caption{
Calculation of the scaled counts in spheres, $\mathcal{D}_2(r)$, as a function of distance, $r$, for 
a pseudo-data sample, that is, a set of $150\,302$ synthetic cosmic objects from a random homogeneous catalogue. 
These pseudo-data catalogue was then compared with other 25 random catalogues in order to obtain $\mathcal{D}_2(r)$. 
As expected, this result shows that the pseudo-data catalogue, i.e., a random set is, indeed, homogeneous.}
\label{fig:D2pseudo}
\end{figure}

The aim of this analysis is to test our whole methodology considering a random catalogue as our data, for this we know refer to it as a {\em pseudo-data} catalogue. 
Our results are displayed in figure~\ref{fig:D2pseudo}. 

In the figure we have $\mathcal{D}_2(r)$ as function of distance $r$ between pairs for a sample of $150\,302$ synthetic cosmic objects from a random catalogue. The random catalogue was used as if it was data and compared with other 25 random catalogues in order to obtain $\mathcal{D}_2(r)$. 
As expected, this result shows that the random set is homogeneous from the smallest scales.

%%%%%%%%%%%%%%%%%%%%%%%%%%%%%%%%%%%%%%%%%%%%%%%%%%

% Don't change these lines
\bsp	% typesetting comment
\label{lastpage}
\end{document}